\documentclass[12pt]{iopart}
\usepackage{amssymb}
\usepackage{amsfonts}
\usepackage{graphicx}
\usepackage{amsthm}
\newcommand{\Ex}{{\rm Ex}}

\newcommand{\C}{{\mathbb C}}
\newcommand{\R}{{\mathbb R}}

\newcommand{\ket}[1]{|{#1}\rangle}
\newcommand{\kb}[1]{|{#1}\rangle \langle{#1}|}

\newcommand{\bra}[1]{\langle{#1}|}
\newcommand{\bkt}[2]{\langle{#1}|{#2}\rangle}

\newcommand{\Ref}[1]{(\ref{#1})}
\newcommand{\wv}[2]{\langle{#1}\rangle_{#2}}
\newcommand{\dwv}[4]{\, _{#1}\langle{#2}\rangle_{#3}^{#4}}

\makeatletter

\newtheorem{fact}{Fact}
\newtheorem*{hardy}{Hardy's Paradox~\cite{HARDY}}

\begin{document}
\title[Strange Weak Values]
	{Strange Weak Values}
 \author{Akio Hosoya$^{1,\dagger}$ and Yutaka Shikano$^{1,2,\ast}$}
 \address{$^1$ Department of Physics, Tokyo Institute of Technology, Meguro, Tokyo 152-8551, Japan}
 \address{$^2$ Department of Mechanical Engineering, Massachusetts Institute of Technology, Cambridge, MA 02139, USA}
 \ead{$^\dagger$\mailto{ahosoya@th.phys.titech.ac.jp}, $^\ast$\mailto{shikano@mit.edu}}
 \date{\today}
\begin{abstract}
We develop a formal theory of the weak values with emphasis on the consistency conditions and a probabilistic interpretation in the counter-factual processes. 
We present the condition for the choice of the post-selected state to give a negative weak value of a given projection operator and strange values of an 
observable in general. The general framework is applied to Hardy's paradox and the spin $1/2$ system
to explicitly address the issues of counter-factuality and strange weak values. The counter-factual arguments which characterize the paradox specifies 
the pre-selected state
and a complete set of the post-selected states clarifies how the strange weak values emerge.
\end{abstract}
\submitto{\JPA}
\maketitle
\section{Introduction}
Many quantum paradoxes appear through attributes which are not actually measured. One may discard all such counter-factuality by 
insisting that attributes can have value only when they are actually measured. A well-known example is Hardy's paradox as we 
discuss later in detail. However, Aharonov and his colleagues have long claimed that there is a consistent and experimentally 
verifiable way to apply the so called {\it weak value} to counter-factual reasoning in quantum mechanics. 
The cost is that we have to allow strange values outside the range of the eigenvalue spectrum of the observable for the weak values. The 
concept of weak value was originally proposed by Aharonov and his collaborators~\cite{AAV,AV91,AR,AV08,AT} in terms 
of the weak measurement combined with pre-selected and post-selected states. Recently, experiments have verified aspects of weak value 
theory both optically~\cite{RSH,PCS,RLS,Pryde,HK,RESCH,Lundeen,Yokota} and in the solid state qubits~\cite{Esteve}. 
On the basis of the weak measurement, a formal theory of the weak value has also been developed~\cite{MJP,JOZSA}~\footnote{A compact 
review of the weak measurement can be seen in these references.}.
However, we believe that the weak values have to be studied on its own right although its measurability is utmost 
important~\footnote{This point of view is strengthened by the recent analysis that demonstrates a reconstruction of 
weak values without using the weak measurements~\cite{JOHA}. There is also a generalization of weak measurement 
using a weak entanglement~\cite{Yokota}.}.
 
The weak value of an observable $A$ at time $t$ is defined by
\begin{equation}
_{\phi}\wv{A}{\psi}^{w}:=\frac{\bra{\phi(t)}A\ket{\psi(t)}}{\bkt{\phi(t)}{\psi(t)}}, \ \bkt{\phi(t)}{\psi(t)} \neq 0, 
\label{defi}
\end{equation}
where the states $\ket{\psi(t)}$ and $\bra{\phi(t)}$ are the solutions of the 
Schr\"{o}dinger equation with given initial and final state conditions, $\ket{\psi(t_i)}=\ket{i}$ and $\bra{\phi(t_f)}=\bra{f}$. 
This time-symmetric approach to quantum mechanics has 
been emphasized in Ref.~\cite{RA}.
Thus, the weak value shares the same spirit as the action principle in the sense
that the both initial and final positions are fixed and can be written in terms of the Feynman path integral (also see Ref.~\cite{Park}). 
 
In this paper we develop a framework of weak value vectors where we fix a pre-selected state and consider a complete orthonormal set 
for post-selected states. We derive the two consistency conditions for the weak value vectors which enables us to interpret them as (complex) 
conditional probabilities for counter-factual processes in quantum mechanics. We shall explicitly give in what cases a weak value becomes strange. 
Two examples are for the demonstration; Hardy's paradox and the spin system. In the former the choice of the pre-selected sate is unique from 
the setting of the paradox and the appearance of the ``negative probability" is due to the counter-factuality. In the latter we discuss the case 
of non-commutative operators to weakly observe.
 
The organization of the present paper is as follows. In Sect.~\ref{Sec_formal} we develop a formal aspect of weak values and give 
the two consistency conditions and a probabilistic interpretation of weak values. In Sect.~\ref{Sec_neg}, we point out that strange values 
result from negative weak values of projection operators, i.e., ``negative probabilities". 
We also show the necessary and sufficient conditions for negative weak values of projection operators and  
give a formula for the most negative weak value. To demonstrate the analysis in Sect.~\ref{Sec_neg}, we examine Hardy's paradox in Sect.~\ref{Sec_Hardy}. 
as an arch-typical case and the spin $1/2$ system in Sect.~\ref{Sec_spin} to see the effect of non-commuting spin operators. 
Section~\ref{Sec_con} is devoted to summary.
\section{Consistency and Probabilistic Interpretation of Weak Values} \label{Sec_formal}
The weak values have been discussed mostly for commuting variables, e.g., optical path projection operators
$\ket{m}\bra{m}\; ( m=1,2, \ldots N )$. Here, the consistency comes from the completeness relation,
\begin{equation}
\sum_{m}{ _{\phi}}\wv{\ket{m}\bra{m}}{\psi}^{w}=1.\;\;\;\;
\mbox{(CONSISTENCY 1)}
\label{consistency1}
\end{equation}
That is, all the possible intermediate states add up to unity in the weak values.
The consistency condition of this kind has been discussed in the literature, e.g.,~\cite{MJP,Vaidman}.
According to the Born rule, which has been verified by numerous experiments, the expectation value of an observable $A$ is given by
\begin{eqnarray}
\bra{\psi}A\ket{\psi}&=\sum_{x}\bkt{\psi}{x}\bra{x}A\ket{\psi} \nonumber \\
& =\sum_{x}|\bkt{\psi}{x}|^2\frac{\bra{x}A\ket{\psi}}{\bkt{x}{\psi}} \nonumber \\
&=\sum_{x}|\bkt{\psi}{x}|^2\;_{x}\wv{A}{\psi}^{w} \nonumber \\
&=\sum _{x} \Pr (x) \, _{x}\wv{A}{\psi}^{w},
\label{Bay}
\end{eqnarray}
provided that the set $\{\bra{x}\}$ does not contain an element such that $\bkt{x}{\psi}=0$. 
Here the $\Pr (x)= |\bkt{\psi}{x}|^2$ is the probability measure to find the state $\ket{x}$ in the pre-selected state $\ket{\psi}$, 
which is defined with respect to a fixed complete set $\ket{x}$ independent of the observable $A$ itself. 
$_{x}\wv{A}{\psi}^{w} := \frac{\bra{x}A\ket{\psi}}{\bkt{x}{\psi}}$ is the weak value of $A$ with $\ket{\psi}$ and $\bra{x}$ being 
the pre-selected and post-selected states, respectively. Comparing Eq. (\ref{Bay}) with the expression for the expectation value $\Ex[A]$ of $A$ in 
the standard probability theory, 
\begin{equation}
\Ex[A]=\sum_{x} \Pr (x) h_{A}(x),
\end{equation}
we see that the statistical average of the weak value becomes the expectation value in quantum mechanics and 
one may interpret the weak value as a complex random variable $h_{A}(x)$~\footnote{The Bayesian-like equation has been pointed out in Ref.~\cite{SH}.};
\begin{equation}
h_{A}(x)= \, _{x}\wv{A}{\psi}^{w} \in \C.
\end{equation}
We emphasize that the probability measure $\Pr (x)= |\bkt{\psi}{x}|^2$ does not depend on the observable $A$.
Aharonov and Botero~\cite[Eq. (2.12)]{AB} derived the formula for the average of observables, which is close to our eq. (\ref{Bay}), though the
observable independence of the probability measure in our presentation seems new.

Further if the operator $A$ is a projection operator $A=\kb{a}$, then the identity (\ref{Bay}) becomes an analog of the
Bayesian rule in  probability theory. The weak value  $_{x}\wv{\kb{a}}{\psi}^{w}$ can be interpreted as the 
conditional probability, which is in general complex, for the process: from the preselected state $\ket{\psi}$ 
to the post selected state $\bra{x}$ via  the intermediate state $\ket{a}$.
The concept of negative probability is not new, e.g., ~\cite{Dirac,FEYNMAN, HOFMANN}. The weak value defined by Eq. (\ref{defi}) is normally 
called the transition amplitude from the state $\ket{\psi}$ to $\bra{\phi}$ via the intermediate state $\ket{a}$ for $A=\kb{a}$, the absolute value 
squared of which is the probability for the process. But the three references quoted above seem to suggest that they might be interpreted as 
probabilities in the case that the process is counter-factual, i.e., the case that the intermediate state $\ket{a} $ is not projectively measured.
The description of intermediate state $\ket{a}$ in the present work is counter-factual or virtual in the sense that the intermediate state would not be observed by projective measurements. 
Feynman's example is the counter-factual ``probability" for an electron to have its spin up in the $x$-direction and also spin down in the $z$-direction ~\cite{FEYNMAN}.

This is very different from the so-called Aharonov-Bergmann-Lebowitz (ABL) formula~\cite{ABL} for the conditional 
probability, which is the modulus squared of the weak value,
\begin{equation}
| _{\phi}\wv{\ket{a}\bra{a}}{\psi}^{w}|^2=\frac{|\bkt{\phi}{a}|^2|\bkt{a}{\psi}|^2}{|\bkt{\phi}{\psi}|^2}=
\frac{\Pr (a; \psi) \Pr (\phi ;a)}{\Pr (\phi; \psi)},
\label{ABL_formula}
\end{equation}
where $\Pr (j ; i) := |\bkt{j}{i}|^2$ is the transitional probability from the state $\ket{i}$ to $\ket{j}$.
The ABL formula has a probabilistic interpretation only if the intermediate state is projectively measured.
For the counter-factual intermediate  state the weak value plays a role of probability rather than amplitude from our view point. 
We will return to this counter-factuality of the weak values in the subsequent sections.
 
The stochastic theory tells us that the variance of the observable $A$ is given by
\begin{eqnarray}
Var(A)&= \sum_{x} |h_{A}(x)|^2 \Pr (x) - \left( \sum _{x}\, h_{A}(x) \Pr (x) \right)^2 \nonumber \\
&=\sum |\, _{x}\wv{A}{\psi}^{w}|^2 \Pr (x) - \left( \sum_{x} \, _{x}\wv{A}{\psi}^{w} \Pr (x) \right)^2 \nonumber \\
&=\bra{\psi}(A-\bra{\psi}A\ket{\psi})^2\ket{\psi}. \label{VAR}
\end{eqnarray}
The last line comes from a straightforward calculation. The result coincides with the conventional ``standard deviation squared" 
in quantum mechanics, which we claim is derived from the identification $h_{A}(x)= \, _{x}\wv{A}{\psi}^{w}$.

Equation (\ref{Bay}) is further generalized to
\begin{equation}
\sum_{x} (\dwv{x}{A}{\psi}{w})^{\ast} |\bkt{x}{\psi}|^2 \dwv{x}{B}{\psi}{w} =\bra{\psi}AB\ket{\psi}.\;\;\;\;\;
\mbox{(CONSISTENCY 2)}
\label{consistency2}
\end{equation}
The proof is similar to that given for Eq. (\ref{Bay}). The formula (\ref{consistency2}) reduces to Eq. (\ref{Bay}) for $B=1$ and for $A=B$ to the $\bra{\psi} A^2 \ket{\psi}$ term
of Eq. (\ref{VAR}) . In particular, if $A$ and $B$ are orthogonal projection operators, the right hand side of Eq. 
(\ref{consistency2}) becomes $\delta_{AB}\bra{\psi}A\ket{\psi}$, which gives an orthonormality condition for the weak values .

Let $A=\ket{a}\bra{a}$ and $B$ be two in general non-commuting projection operators. If we adopt the previous interpretation of weak values 
as a conditional complex probability, the left hand side of Eq. (\ref{consistency2}) implies that the right hand side $\bra{\psi}AB\ket{\psi}$ 
has a meaning of the joint probability for $B$ and then $A$ to occur if it is real, i.e., $A$ and $B$ are commutable in the sense of 
expectation value~\footnote{The authors thank Professor Ozawa for this comment who has formulated the state-dependent joint probability~\cite{OZAWA}.}. 
Putting the other way, we confirm this by the identity,
\begin{equation}
\bra{\psi}AB\ket{\psi}=|\bkt{a}{\psi}|^2\frac{\bra{a}B\ket{\psi}}{\bkt{a}{\psi}}=|\bkt{a}{\psi}|^2\dwv{a}{B}{\psi}{w}.
\label{joint}
\end{equation}
Steinberg~\cite{Steinberg1,Steinberg2} addressed the issue of the tunneling traversal time from the view point of the weak measurement in which he proposed
the joint and conditional probabilities for the counter-factual cases in quantum mechanics. We regard the relation (\ref{joint}) of the joint and conditional
probabilities as consistency conditions for a set of weak values in a framework where the weak value is promoted to a fundamental concept
independent of the weak measurement model.

It is interesting to point out that the quantity $\bra{\psi}(A-B)^2\ket{\psi}$, which is an indicator of the difference of 
the two quantities $A$ and $B$ in the state  $\ket{\psi}$, can be evaluated by the experimentally accessible weak values via 
the following formula:
\begin{equation}
\sum_{x}{ (\, _{x}\wv{A}{\psi}^{w}- \, _{x}\wv{B}{\psi}^{w})^{*}}|\bkt{x}{\psi}|^2 
(\, _{x}\wv{A}{\psi}^{w}- \, _{x}\wv{B}{\psi}^{w}) =\bra{\psi}(A-B)^2\ket{\psi}. \;\;\;\;\;
\label{consistency3}
\end{equation}
With $\ket{\psi}$ being the pre-selected state, the probability to get a post-selected state $\ket{x}$ is $\Pr (x)=|\bkt{x}{\psi}|^2$ 
so that the left hand side can be expressed as 
\begin{equation}
\sum_{x} \Pr (x)|h_{A}(x)-h_{B}(x)|^2.
\end{equation}
which can be computed exploiting all the data exhausting all possible post-selected states. 
In terms of the weak measurements, a particular post-selection $x$ shifts the
center of the distribution of the pointer position by $h_{A}(x)$ when $A$ is measured and by $h_{B}(x)$ for $B$. The difference  $h_{A}(x)-h_{B}(x)$ varies
depending on the post-selected state $\ket{x}$. The weighted sum of the square of the difference over the post-selected states 
gives the quantity $\bra{\psi}(A-B)^2\ket{\psi}$.
Note that the probability measure defined independently of the observed quantities helps to give a symmetric expression for the two observables $A$ and $B$.
We would like to remark that the following simple fact.
\begin{fact}
Fix the quantum state $\ket{\psi}$. The observables $A$ and $B$ are equivalent for the quantum state $\ket{\psi}$ if and only if 
\begin{equation}
	\bra{\psi} (A - B)^2 \ket{\psi} = 0.
	\label{equiv_eq}
\end{equation}
\label{equiv}
\end{fact}
Note that this condition implies that
\begin{equation}
	\,_{\phi} \wv{(A - B)}{\psi}^{w}=0
	\label{weak_eq}
\end{equation}
for any post-selected states $\bra{\phi}$. 
That is, two observables are equivalent if their weak value vectors are identical.
This fact will be used to
check whether the state $\ket{\psi}$ is the correct one for an 
experimental setup in Sect. \ref{Sec_Hardy_Pre}.
 
A keen reader may notice that  only the real part of the weak value is relevant in the Bayesian-like rule \Ref{Bay} so that there is an option 
that we interpret only the real part as a conditional probability. Then it would be interesting to consider the restricted cases where 
the weak values are real. This also demands the reality of the joint probability  \Ref{joint}, which implies the commutativity of 
the observables $A$ and $B$ relative to the pre-selected state $\ket{\psi}$, i.e., there is no Heisenberg-Robertson uncertainty 
$\Delta A\Delta B\geq 0$ for the state $\ket{\psi}$, where $\Delta A$ and $\Delta B$ are the square root of the variances of $A$ and $B$. 
A simple example is the case: $A=\sigma_x, B=\sigma_y$ and $\ket{\psi}=(\ket{0}+\ket{1})/\sqrt{2}$ which gives $\Delta \sigma_x 
\Delta \sigma_y = |\bra{\psi} \sigma_z \ket{\psi}| =0$. Namely, there is no uncertainty bound for the $\sigma_x$ and the $\sigma_y$ 
because they are commuting in the state $\ket{\psi}$. Hereafter we will be mainly 
concerned with the real weak value unless otherwise stated.
\section{Condition for a Strange Weak Value} \label{Sec_neg}
When some of the weak values of the projection operators exceed unity, at least one of the other weak values, say 
$\Pr_m:=\dwv{f}{\kb{m}}{i}{w}$ becomes negative. One might interpret the $\Pr_m$ as a ``negative" conditional   
probability for the path $\ket{i}\rightarrow \ket{m}\rightarrow \ket{f}$. 
We can see the negative conditional probability is also
responsible for the various ``strange" weak values, e.g., the weak value of the $z-$component of the spin of a Dirac particle 
can be very large, say $100 \hbar$~\cite{AAV}~\footnote{Of course, this is not mathematically strange at all. It is strange only if 
one insists on the probabilistic interpretation of the weak values.}.
Let the spectral decomposition of an observable $A$ be
\begin{equation}
A=\sum_{n=1}^{N}a_{n}\ket{n}\bra{n},
\end{equation}
where the eigenvalues $a_{n}$ obey the order relation $a_{N}> a_{N-1}>\cdots >a_{1}$ and $\ket{n}$ is the eigenstate with the eigenvalue $a_n$.
We concentrate on the maximum and minimum states, $\ket{N}$ and $\ket{1}$. The pre-selected state is assumed to be 
\begin{equation}
\ket{\psi}=\alpha\ket{N}+\beta\ket{1}, \;\;\alpha, \beta \in\R
\end{equation}
Choose the post-selected states $\ket{\phi}$ as
\begin{equation}
\ket{\phi}=\frac{\ket{N}+\ket{1}}{\sqrt{2}}
\end{equation}
to obtain the weak value of $A$ as 
\begin{equation}
_{\phi}\wv{A}{\psi}^w ={a_{N}} \, _{\phi}\wv{\ket{N}\bra{N}}{\psi}^w +{ a_{1}} \, _{\phi}\wv{\ket{1}\bra{1}}{\psi}^w = 
a_{N} + (a_{1}-a_{N}) \, _{\phi}\wv{\ket{1}\bra{1}}{\psi}^w.
\end{equation}
This can be larger than the maximum eigenvalue $a_{N}$ of $A$ for negative $_{\phi}\wv{\ket{1}\bra{1}}{\psi}^w$, i.e., for $\alpha/\beta<-1$. 
Similarly, we can analyze the case that the weak value is smaller than the minimum eigenvalue. The point is that at least one of the weak 
values of the projection operators is negative to have a weak value outside the range of the spectrum. The converse also holds, i.e., 
if the weak values of all the projection operators are positive, the weak value of the observable $A$ is not strange.

In the weak measurement the  post-selection biases the original probability distribution to push the mean value of the pointer position
sometimes to a strange value outside the spectrum. There should be the optimal post-selection which shifts the pointer position most.
We are going to find the condition for a weak value to be strange. From the argument stated above, we see that it is sufficient to
consider one of the eigenvectors of the observable $\ket{n}$ and find the condition for the weak value of the projection operator $\ket{n}\bra{n}$
to be negative.

Let us consider the weak value of a projection operator $\ket{n}\bra{n}$,
\begin{equation}
	w(\phi):= \,_{\phi}\wv{\ket{n}\bra{n}}{\psi}^{w} = \frac{\bkt{\phi}{n}\bkt{n}{\psi}}{\bkt{\phi}{\psi}}.
\end{equation}
We are going to find the optimal
post-selection to obtain the most negative weak value $w(\phi)$ for a given pre-selected state and the projection operator 
$\ket{n}\bra{n}$ to be weakly measured. Note that the probability $\Pr (\phi) :=|\bkt{\phi}{\psi}|^2$ to obtain $\ket{\phi}$ has to be 
finite for the weak value to be finite. Otherwise, the weak measurement would produce null results. The smaller the probability 
$Pr (\phi)$ is, the more we need to repeat the weak measurements. We fix the value $\Pr (\phi) =|\bkt{\phi}{\psi}|^2 =\cos^2\xi$ with $\xi \in (0, \pi/2)$
~\footnote{However, there is a proposal of rather sophisticated single-shot weak measurement, which is out of scope of the present work~\cite{TOL}.}.

To simplify the argument we also assume that all the inner products are real so that we can visualize the state vectors in the multi-dimensional 
real vector space. In particular, the post-selected state $\ket{\phi}$ is lying on the cone defined by $\bkt{\phi}{\psi}=\cos\xi$. 
Let the angle between $\ket{n}$ and $\ket{\psi}$ be $\theta_{n} \in (0, \pi/2)$.
Then the most negative weak value $w(\phi)$ is given by a state $\ket{\phi}$ which is the intersection of the cone $\bkt{\phi}{\psi}=\cos\xi$ and the
plane defined by the two vectors $\ket{\psi}$ and $\ket{n}$. Then the weak value $w(\phi)$ becomes
\begin{equation}
	w(\phi) = \frac{\bkt{\phi}{n}\bkt{n}{\psi}}{\bkt{\phi}{\psi}}=\frac{\cos(\theta_{n}+\xi)\cos\theta_{n}}{\cos\xi}.
\label{mostnegative}
\end{equation}
This can be negative if the angle between $\ket{n}$ and $\ket{\phi}$, $\theta_{n}+\xi$ is obtuse, i.e.,
\begin{equation}
	\xi +\theta_{n}> \pi/2,
\end{equation}
provided that the angles $\xi, \theta_n$ are acute, which can be assumed without loss of generality. Note that the above condition cannot be met if
the angle $\xi$ is too small, i.e., the probability $\Pr (\phi)=\cos^2\xi$ is too close to unity. There is a trade-off between the labor and the fruit, 
i.e., the repetition of the weak measurements and the large negative weak value.

We turn to an interesting but more technical problem to find the most strange weak value of the observable $A$,
\begin{equation}
_{\phi}\wv{A}{\psi}^{w} = \frac{{\sum_{n=1}^{N}a_{n}\bkt{\phi}{n}}\bkt{n}{\psi}}{\bkt{\phi}{\psi}},
\end{equation}
under the constraints: $\bkt{\phi}{\psi}=\cos\xi, \bkt{\phi}{\phi}=1$. Introducing the corresponding Lagrange multipliers 
$\lambda$ and $\mu$, and following the standard variation method, we obtain the equation for $\bkt{n}{\phi}$ as
\begin{equation}
\sum_{n=1}^{N}a_{n}\ket{n}\frac{\bkt{n}{\psi}}{\cos\xi}-\lambda\ket{\psi}-\mu\ket{\phi}=0.
\label{optimalphi}
\end{equation}
The formal solution is
\begin{equation}
\ket{\phi}=\frac{1}{\mu} \left[ \sum_{n=1}^{N}a_{n}\ket{n}\frac{\bkt{n}{\psi}}{\cos\xi}-\lambda\ket{\psi} \right],
\label{optimalphi2}
\end{equation}
where the multipliers $\lambda$ and $\mu$ are given by
\begin{eqnarray}
\lambda&=\sum_{n=1}^{N}\frac{a_{n}}{\sin^2\xi} \left[ \frac{\bkt{n}{\psi}^2}{\cos\xi}-\bkt{\phi}{n}\bkt{n}{\psi} \right], \\
\mu&=-\sum_{n=1}^{N}\frac{a_{n}}{\sin^2\xi} \left[ \bkt{n}{\psi}^2-\frac{\bkt{\phi}{n}\bkt{n}{\psi}}{\cos\xi} \right].
\label{lag}
\end{eqnarray}
The post-selection by the state (\ref{optimalphi2}) gives the most strange weak value. By projecting out this state 
by the base $\ket{n}$ we have a closed set of algebraic equations
for $\bkt{\phi}{n}$:
\begin{eqnarray}
&a_{n}\bkt{\phi}{n}\sin^2\xi \nonumber \\
&=\bkt{\phi}{n} \left[ \sum_{m=1}^{N}a_{m}(\bkt{\phi}{m}-\bkt{\psi}{m}\cos\xi)\bkt{\phi}{m} \right] \nonumber \\
& \ \ \ \ - \bkt{\psi}{n} \left[ \sum_{m=1}^{N}a_{m}(\bkt{\phi}{m}\cos\xi -\bkt{\psi}{m}) \bkt{\phi}{m} \right] \ \ (n=1\dots N).
\end{eqnarray}
For the particular case of single state  $\ket{n}$, the previous result is reproduced by
one of the solutions for $\bkt{\phi}{n}=\cos(\theta+\xi)$.

We are going to demonstrate the above statement in two explicit examples;
Hardy's setting and the spin $1/2$ system. In the first example, the weak values are real and  the ``negative probability" appears, 
while in the second spin system complex weak values appear and the spin $x$-component becomes large, say, $100\hbar$. 
\section{Hardy's Paradox} \label{Sec_Hardy}
\begin{figure}[t]
	\begin{center}
	\includegraphics[width=10cm]{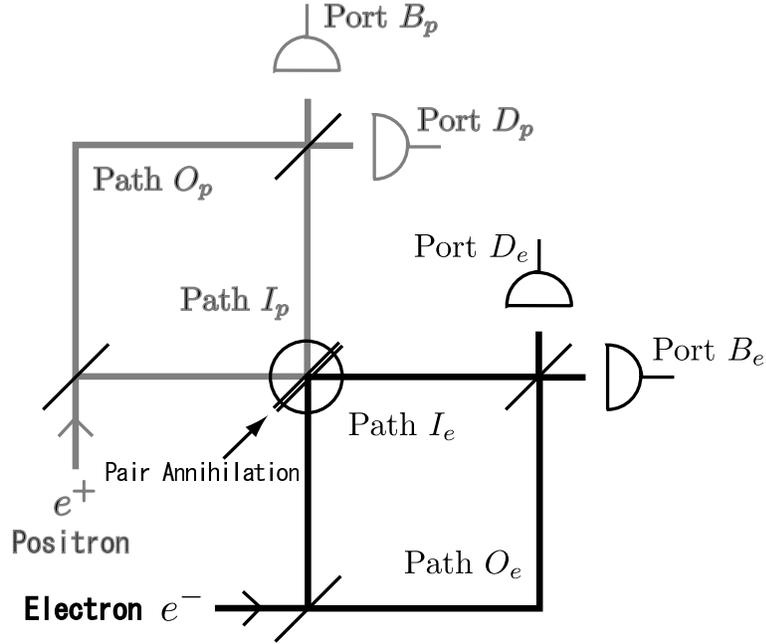}
	\caption{Hardy's setup: The Hardy paradox is explained in the main text.}
	\label{fig}
	\end{center}
\end{figure}

Let us recall what Hardy's paradox is. 
\begin{hardy}
The two Mach-Zehnder interferometers for an electron and a positron are combined in such a way that one of the arms, say, $I_{e}$ of the interferometer 
for the electron is crossed with one of the arms $I_p$ for the positron. If the electron and the positron meet at the crossing they are supposed to annihilate each other. 
The setting is illustrated in Fig.~\ref{fig}.
\begin{description}
\item[1)] 
	Both the electron and positron cannot be in the inner path $I_e$ and $I_p$ at the same time since 
	they would annihilate each other. Therefore, either electron or positron is in the outer path $O_e$ or $O_p$. \label{hardy1}
\item[2)]
	Let the positron be in the path $O_p$ so that the the electron is not affected by the positron. The Mach-Zehnder interferometer
	works as usual; the intermediate state is $O_p(I_e+O_e)$ (a shorthand for $\ket{O_p}\ket{\frac{(I_e+O_e)}{\sqrt{2}}}$ which we use 
	in this section for notational simplicity) and therefore the electron registers a click at detector $B_e$, while the positron clicks either the port $B_p$ or $D_p$. Similarly, for the electron in $O_e$, the intermediate
	state will be $(I_p+O_p)O_e$ and therefore the positron registers a click at detector $B_p$, while the electron clicks either $B_e$ or $D_e$.
	In either case, only possible cases are: $B_pB_e, D_pB_e, B_pD_e$, but $D_pD_e$ is not. \label{hardy2}
\item[3)] However, quantum mechanics tells us that $D_pD_e$ clicks with the probability $1/12$, which has also been verified by experiments. \label{hardy3}
\end{description}
\end{hardy}
The line of reasoning 2) is not a consequence of measurements but an inference by the counter-factual argument~\cite{UNRUH}.
For example, if the positron is found in $O_p$ by measurement, the electron is definitely in $(I_e+O_e)$ ending up with $B_e$, which can be 
demonstrated  as an actual experiment.
Another different experiment would be the one starting from the fact that the electron found in $O_e$. But one cannot
infer anything about the non-measured quantity by combining  the two different experiments of non-commuting observables $O_p(I_e+O_e)$ and $(I_p+O_p)O_e$. 
The most obvious way to resolve Hardy's paradox is to dismiss the counter-factual reasoning all together. 

However, Aharonov and his colleagues have long advocated the use of weak values which are experimentally accessible to provide experimental credence to 
the counter-factual arguments~\cite{hardy_aharonov}. The weak value gives a consistent description of the quantum
evolution of the intermediate states at the cost of strange values, i.e., the value outside the range of the spectrum of observable. Actually they 
have been recently demonstrated in the optical settings~\cite{Lundeen,Yokota}. 
It should be noted that the Hardy paradox has also been analyzed in the different contexts~\cite{Stapp,Mermin,Griffiths,McCall,Sokolovski,Tsang}.
We are going to closely examine the counter-factual step 2) 
by looking at the corresponding weak values. We will see what is missing in the argument of 2) and how the counter-factual argument works in terms 
of weak values.
\subsection{Pre-selected State of the Hardy Paradox} \label{Sec_Hardy_Pre}
In this section, we justify the standard initial quantum state, 
\begin{equation}
	\ket{\psi} = \frac{1}{\sqrt{3}} ( \ket{I_pO_e} + \ket{O_pI_e} + \ket{O_pO_e} ),
	\label{originalHardy}
\end{equation}
on the basis of the counter-factual arguments 2) in the original Hardy paradox 
and then show how to experimentally check whether the initial quantum state is 
the right hand side of Eq. (\ref{originalHardy}) or not by evaluating the weak values.

The pre-selected state is in general given by 
\begin{equation}
	\ket{\psi} = \eta \ket{I_pI_e} + x \ket{I_pO_e} + y \ket{O_pI_e} + z \ket{O_pO_e}
\end{equation}
with the coefficients $\eta, x, y, z \in \mathbb{C}$ satisfying the normalization condition, $|\eta|^2 + |x|^2 + |y|^2 + |z|^2 = 1$. 
We are going to determine the coefficients by the following counter-factual arguments. Since  the electron going through the path $O_e$ 
does not affect the path of the positron, the projection operators $P[O_p(I_e+O_e)]$ and $P[O_p \otimes id.]$ are equivalent in the initial 
quantum state $\ket{\psi}$. Here $id.$ denotes the identity operator. Similarly, $P[(I_p+O_p)O_e]$ and $P[id. \otimes O_e]$ are equivalent in 
$\ket{\psi}$. Furthermore, since the electron and the positron simultaneously going through the paths $I_p$ and $I_e$ would 
annihilate each other, $P[I_pO_e]$ and $P[I_p \otimes id.]$ and $P[O_pI_e]$ and $P[id. \otimes I_e]$ are also equivalent for $\ket{\psi}$. 
According to Fact~\ref{equiv}, the above counter-factual arguments demand
\begin{eqnarray}
	& \bra{\psi} (P[O_p(I_e+O_e)] - P[O_e \otimes id.])^2 \ket{\psi} \nonumber \\ & \ \ \ \ \ \ = \bra{\psi} (P[O_p(I_e+O_e)] - P[O_p \otimes id.]) 
	\ket{\psi} = 0 \label{eq1}\\
	& \bra{\psi} (P[(I_p+O_p)O_e] - P[id. \otimes O_e])^2 \ket{\psi} \nonumber \\ & \ \ \ \ \ \ = \bra{\psi} (P[(I_p+O_p)O_e] - P[id. \otimes O_e]) 
	\ket{\psi} = 0 \label{eq2}\\
	& \bra{\psi} (P[I_pO_e] - P[I_p \otimes id.])^2 \ket{\psi} \nonumber \\ & \ \ \ \ \ \ = \bra{\psi} (P[I_pO_e] - P[I_p \otimes id.]) 
	\ket{\psi} = 0 \label{eq3} \\
	& \bra{\psi} (P[O_pI_e] - P[id. \otimes I_e])^2 \ket{\psi} \nonumber \\ & \ \ \ \ \ \ = \bra{\psi} (P[O_pI_e] - P[id. \otimes I_e]) \ket{\psi} = 0 \label{eq4}
\end{eqnarray}
From Eqs. (\ref{eq3}) or (\ref{eq4}), we obtain $\eta = 0$. From Eqs. (\ref{eq1}) and (\ref{eq2}), we obtain $x= y = z$. Taking into account the 
normalization condition, we obtain the standard Hardy initial state (\ref{originalHardy}). The second equations of Eqs. (\ref{eq1}),  (\ref{eq2}), 
(\ref{eq3}) and  (\ref{eq4}) also imply that the initial quantum state can be experimentally tested by the joint expectation value of 
the paths for the electron and the positron. 

Furthermore, we will show that this initial state can also be verified by the joint weak measurement, that is, the state can be 
constructed from a set of weak values. The inner product of the pre- and post-selected states are 
\begin{eqnarray}
	\zeta_{D_pD_e} & := \bkt{D_pD_e}{\psi} = \frac{1}{2}(\eta - x - y + z), \\ 
	\zeta_{D_pB_e} & := \bkt{D_pB_e}{\psi} = \frac{1}{2}(\eta + x - y - z), \\ 
	\zeta_{B_pD_e} & := \bkt{B_pD_e}{\psi} = \frac{1}{2}(\eta - x + y - z), \\ 
	\zeta_{B_pB_e} & := \bkt{B_pB_e}{\psi} = \frac{1}{2}(\eta + x + y + z).
\end{eqnarray}
We obtain the following table of weak values for the pre-selected state $\ket{\psi}$ and the post-selected state $\ket{\phi}$:
\begin{equation}
\begin{array}{c|cccc}
{\rm Weight} & \zeta_{D_pD_e}^2 & \zeta_{D_pB_e}^2 & \zeta_{B_pD_e}^2 & \zeta_{B_pB_e}^2 \\
\phi & D_pD_e & D_pB_e & B_pD_e & B_pB_e \\ \hline
O_p(I_e+O_e) & 0 & - \frac{y + z}{\zeta_{D_pB_e}} & 0 & \frac{y + z}{\zeta_{B_pB_e}} \\  
(I_p+O_p)O_e & 0 & 0 & -\frac{x + z}{\zeta_{B_pD_e}} & \frac{x + z}{\zeta_{B_pB_e}} \\  
I_pO_e & - \frac{x}{2 \zeta_{D_pD_e}} & \frac{x}{2 \zeta_{D_pB_e}} & - \frac{x}{2 \zeta_{B_pD_e}} & \frac{x}{2 \zeta_{B_pB_e}} \\  
O_pI_e & - \frac{y}{2 \zeta_{D_pD_e}} & - \frac{y}{2 \zeta_{D_pB_e}} & \frac{y}{2 \zeta_{B_pD_e}} & \frac{y}{2 \zeta_{B_pB_e}} \\
O_p \otimes id. & \frac{- y + z}{2 \zeta_{D_pD_e}} & \frac{- y - z}{2 \zeta_{D_pB_e}} & \frac{y - z}{2 \zeta_{B_pD_e}} & \frac{y + z}{2 \zeta_{B_pB_e}} \\  
I_p \otimes id. & \frac{\eta - x}{2 \zeta_{D_pD_e}} & \frac{\eta + x}{2 \zeta_{D_pB_e}} & \frac{\eta - x}{2 \zeta_{B_pD_e}} & \frac{\eta + x}{2 \zeta_{B_pB_e}} \\  
id. \otimes O_e & \frac{- x + z}{2 \zeta_{D_pD_e}} & \frac{x - z}{2 \zeta_{D_pB_e}} & \frac{- x - z}{2 \zeta_{B_pD_e}} & \frac{x + z}{2 \zeta_{B_pB_e}} \\
id. \otimes I_e & \frac{\eta - y}{2 \zeta_{D_pD_e}} & \frac{\eta - y}{2 \zeta_{D_pB_e}} & \frac{\eta + y}{2 \zeta_{B_pD_e}} & \frac{\eta + y}{2 \zeta_{B_pB_e}}  
\end{array},
\label{table1}
\end{equation}
where the ``Weight" is the probability of transition from the pre-selected state $\ket{\psi}$ to the post-selected state $\ket{\phi}$. 
Here, the column of the post-selected state $\ket{\phi}$ shows the weak values for each observable, e.g. $O_p(I_e+O_e)$.
Interestingly our table (\ref{table1}) is similar to the one in Feynman's paper~\cite{FEYNMAN}.
From the same counter-factual arguments as before together with Eq. (\ref{weak_eq}), we see that
\begin{eqnarray}
	InPositron(D_pD_e) &:= \, _{D_pD_e} \wv{I_pO_e}{\psi}^{w} - \, _{D_pD_e} \wv{I_p \otimes id.}{\psi}^{w}=0, \label{inp}\\
	InEletron(D_pD_e) & := \, _{D_pD_e} \wv{O_pI_e}{\psi}^{w} - \, _{D_pD_e} \wv{id. \otimes I_e}{\psi}^{w}=0, \label{ine}\\
	OutPositron(D_pD_e) &:= \, _{D_pD_e} \wv{O_p(I_e + O_e)}{\psi}^{w} - \, _{D_pD_e} \wv{O_p \otimes id.}{\psi}^{w}=0, \label{outp} \\
	OutEletron(D_pD_e)& := \, _{D_pD_e} \wv{(I_p + O_p)O_e}{\psi}^{w} - \, _{D_pD_e} \wv{id. \otimes O_e}{\psi}^{w}=0. \label{oute}
\end{eqnarray}
Using the results of the above table in Eqs. (\ref{inp}) and (\ref{ine}) yields 
\begin{equation}
	\eta - x = - x \ \ \ {\rm or} \ \ \ \eta - y = - y
\end{equation}
to obtain that the coefficient of $\ket{I_pI_e}$ must be $\eta = 0$. 
Similarly from Eqs. (\ref{outp}) and (\ref{oute}), we have
\begin{equation}
	- y + z = 0 \ \ \ {\rm and} \ \ \ - x + z = 0
\end{equation}
to obtain the coefficients $x = y = z = 1 / \sqrt{3}$ by the normalization condition. 

We therefore conclude that
\begin{fact}
The following conditions
\begin{enumerate}
	\item $InPositron(D_pD_e) = InElectron(D_pD_e) = 0$,
	\item $OutPositron(D_pD_e)= OutElectron(D_pD_e) = 0$,
\end{enumerate}
are satisfied if and only if the pre-selected state is Eq. (\ref{originalHardy}).
\end{fact}
Note that we only need a particular post-selected state $\ket{D_pD_e}$ to verify whether a prepared setup 
gives the standard Hardy's initial state (\ref{originalHardy}) or not. 

It is interesting to point out that the relations between the weak values can be used to evaluate the weak values of the 
non-local operators like $I_pO_e, O_pI_e, O_p(I_e+O_e), (I_p+O_p)O_e$ using those of the local operators $I_p \otimes id., id. \otimes I_e, 
O_p \otimes id., id. \otimes O_e$, respectively, provided that the pre-selected state is Eq. (\ref{originalHardy}). 
\subsection{Weak values in the Hardy Paradox}
For a given pre-selected state $\ket{\psi}$ and a post-selected state $\bra{\phi}$ the weak value of an observable $A$ is defined by
\begin{equation}
_{\phi }\wv{A}{\psi }^{w}= \frac{\bra{\phi}A\ket{\psi}}{\bkt{\phi}{\psi}}.
\end{equation}
This can be considered as a vector with component index $\phi$ while the preselected state $\ket{\psi}$ is fixed.
This vector notation is convenient to find the condition for the strange weak values as we will see below.
The pre-selected state in Hardy's paradox is 
\begin{equation}
\ket{\psi}=\frac{\ket{I_pO_e+O_pI_e+O_pO_e}}{\sqrt{3}}.
\end{equation}
as we have discussed in the previous subsection. The possible post-selected states are
\begin{equation}
\ket{D_pD_e}, \ket{D_pB_e}, \ket{B_pD_e}, \ket{B_pB_e}.
\end{equation}
If we choose a set of non-commuting projection operators:
\begin{equation}
P[O_p(O_e+I_e)], P[(O_p+I_p)O_e], P[O_pI_e+I_pO_e],
\end{equation}
then, the weak values for them are real and summarized in following table: 
\begin{equation}
\begin{array}{c|ccccc}
{\rm Weight} & 1/12 & 1/12 & 1/12 & 3/4 & \\
\phi & D_pD_e & D_pB_e & B_pD_e & B_pB_e & {\rm Average} \\ \hline
O_p(I_e+O_e) & 0 & 2 & 0 & 2/3 & 2/3 \\  
(I_p+O_p)O_e & 0 & 0 & 2 & 2/3 & 2/3 \\ 
I_pO_e+O_pI_e & 2 & 0 & 0 & 2/3 & 2/3
\end{array},
\label{table2}
\end{equation}
where ``Average" is defined as the sum of the weak value times ``Weight" over the post-selected states.
Note that all the components of the weak values of the non-commuting projection operators $P[I_pO_e+O_pI_e], P[O_p(I_e+O_e)], P[(I_p+O_p)O_e]$ are positive.
The first two cases, $O_p(I_e+O_e)$ and $(I_p+O_p)O_e$ have played important roles in the previous  counter-factual arguments. The non-commutativity 
is explicit as the non-orthogonality of the vectors, e.g., $(1, 0, 0, 1/3)$ and $(0, 1, 0,1/3)$. The weight is defined by the square of the inner product as  
$1/12, 1/12, 1/12$ and $3/4$. Furthermore, the average means the weighted average of the weak value for each row. 
It is interesting to point out that if $D_e$ and $D_p$ click the intermediate state should be definitely an 
entangled state $I_pO_e+O_pI_e$. If one tries to include the path $O_pI_e+I_pO_e$, the argument would become more counter-factual, because any local measurement 
would destroy this entangled  state.
This nonlocal state is one of the origin of Hardy's paradox parallel to the Einstein-Podolsky-Rosen (EPR) paradox~\cite{EPR}, because the classical 
argument 2) excludes this possibility of entanglement. These arguments are counter-factual in the sense that the identity of the intermediate states cannot 
be revealed by projective measurements. However, the weak measurement can do the job.

One might be curious about the strange value $2$, e.g., for the projection operator $O_p(I_e+O_e)$ and the post-selected state $B_pD_e$ in Table (\ref{table2}). 
This can be explained in a counter-factual way by looking at the  projection operators orthogonal to $O_p(I_e+O_e)$. For example, take $I_pO_e$ 
orthogonal to $O_p(I_e+O_e)$. The table becomes
\begin{equation}
\begin{array}{c|ccccc}
{\rm Weight} & 1/12 & 1/12 & 1/12 & 3/4 & \\
\phi & D_pD_e & D_pB_e & B_pD_e & B_pB_e & {\rm Average} \\ \hline
O_p(I_e+O_e) & 0 & 2 & 0 & 2/3 & 2/3 \\
I_pO_e & 1 & -1 & 1 & 1/3 & 1/3 \\ \hline
{\rm Sum} & 1 & 1 & 1 & 1 & 1
\end{array}.
\label{table3}
\end{equation}
We immediately see that the weak value $-1$ for $I_pO_e$ with $D_pB_e$ partially cancels $2$ for $O_p(I_e+O_e)$ to give $1$ as the sum. Note that the other
orthogonal projection operator to $O_p(I_e-O_e)$ gives the zero weak values supporting the counter factual argument 2) of Hardy's paradox.
Obviously, we can make the completely parallel argument for $(I_p+O_p)O_e$ with the post-selected state $B_pD_e$. 

It is interesting to look at the factual cases when we measure the intermediate states corresponding to the weak values in Table (\ref{table3}).
Suppose we locally measure the intermediate state of the positron to find $O_p$ implying the electron state is $\ket{I_e+O_e}/\sqrt{2}$, 
the probability of which is $2/3$ according to the Born rule $|\bkt{\frac{O_p(I_e+O_e)}{\sqrt{2}}}{\psi}|^2$. Either positron detector $B_p$ or $D_p$ 
clicks with probability $1/2$ and only $B_e$ of the electron detectors clicks. Thus, the probabilities are $0$ for 
$D_pD_e$ and $B_pD_e$, while $(1/2)\times(2/3)=1/3$ for $D_pB_e$ and $B_pB_e$. Similarly, in the case when $I_p$ is the positron intermediate state, the 
probabilities are $(1/4)\times (1/3)=1/12$ for all $D_pD_e, D_pB_e, B_pD_e$ and $B_pB_e$. These factual results can be also reproduced by the ABL 
formula~\cite{ABL}, i.e., the absolute value squared of the weak values multiplied by the weight of the post-selected state. In the case of $O_p$ for the positron, 
the ABL formula gives $0, 2^2\times(1/12)=1/3, 0$ and $(2/3)^2\times(3/4)= 1/3$ for $D_pD_e, D_pB_e, B_pD_e$ and $B_pB_e$, respectively. In the case of $I_p$, it reads 
$1^2\times(1/12)=1/12,(-1)^2\times(1/12)=1/12,1^2\times(1/12)=1/12$ and $(1/3)^2\times(3/4)=1/12$ for $D_pD_e, D_pB_e, B_pD_e$ and $B_pB_e$, respectively. Note that
this coincidence is possible only for the factual and mutually orthogonal set of projection operators and does not work for the counter-factual cases. 
In the latter case we have to resort to fully quantum mechanical computations rather than the semi-classical counter-factual arguments. 
For the factual cases, that is, when we projectively measure the intermediate state $\ket{n}$, we can confirm the ABL formula for the 
partial probabilities by Eq. (\ref{ABL_formula}).

So far we have fixed the post-selected states as $D_pD_e, D_pB_e, B_pD_e$ and $B_pB_e$. 
Here we consider a different choice of post-selected state $\ket{\phi}$ keeping the same pre-selected state 
$ \ket{\psi} = ( \ket{I_pO_e} + \ket{O_pI_e} + \ket{O_pO_e} )/\sqrt{3}$ (\ref{originalHardy}).
For the intermediate $\ket{I_pO_e}$, the most negative weak value will be given
by a post-selected state which lies in the plane spanned by $\ket{\psi}$ and $\ket{n}$ as we noted in 
Sect. \ref{Sec_neg}. We parametrize it as
$\ket{\phi}:=\cos{\theta}\ket{I_pO_e}+\sin{\theta}\frac{\ket{O_pI_e+O_pO_e}}{2}$ by an angle $\theta$. 
It can be shown by a simple calculation that the weak value $ w(\phi) = \frac{\bkt{\phi}{n}\bkt{n}{\psi}}{\bkt{\phi}{\psi}}
=\cos(\theta_{n}+\xi)\cos\theta_{n} / \cos\xi$ (\ref{mostnegative})
can be $-\infty$ for $\xi=\pi/2$ and therefore $\theta=-\pi/4$, since $\bkt{\phi}{\psi}=\cos\xi= (\sin\theta+\cos\theta) / \sqrt{3}$.
 
On the other hand, the weak values for the orthogonal projection operators are
\begin{equation}
\begin{array}{c|ccccc}
{\rm Weight} & 1/12 & 1/12 & 1/12 & 3/4 & \\
\phi & D_pD_e & D_pB_e & B_pD_e & B_pB_e & {\rm Average} \\ \hline
I_pO_e & 1 & -1 & 1 & 1/3 & 1/3 \\
O_pI_e & 1 & 1 & -1 & 1/3 & 1/3 \\
O_pO_e & -1 & 1 & 1 & 1/3 & 1/3 \\ \hline
{\rm Sum} & 1 & 1 & 1 & 1 & 1
\end{array}.
\label{nibu}
\end{equation}
Let us check the consistency conditions 1 and 2 in the Table (\ref{nibu}).
The last row indicates that the consistency condition 1 (\ref{consistency1})
is satisfied, since the sum of the weak values over the complete set of orthonormal projectors is unity for all the post-selected states. 
To confirm the consistency 2 (\ref{consistency2}) we see in the rightmost column that the weighted average
of the weak values over the post-selected states for each intermediate state
becomes the probability for the intermediate state to be actually observed
by projective measurements.
 Let us look at the column of $D_pD_e$. The apparent puzzle is that
the ``probabilities" for $I_pO_e$ or $O_pI_e$ add up to $2$ exceeding $1$.
The weak value resolution to it is that the ``probability" of the remaining $O_pO_e$
is $-1$ so that the total ``probability" is $1+1-1=1$!

With this example we can see more clearly how the minus sign emerges in the table of the weak values. 
Suppose we have two non-orthogonal projection operators like $I_pO_e+O_pI_e$ and $O_p(I_e+O_e)$. 
Assume that all the components of them are positive. Otherwise, we have done. Keep one of them as it is 
and find an orthogonal vector to it by the Schmidt orthogonalization of the other one. One may be easily convinced that the new one 
contains at least one negative component, unless the new one is
lying along one of the axis. If the latter is the case we can go on picking up the third one and then Schmidt diagonalization. 
If we end up with the configuration of the weak values which are
all along the different axes, this is against our assumption of counter-factuality. That is, we could  verify the weak value by 
a projective measurement. In Hardy's paradox, the counter-factuality is unavoidable so that the negative weak value should appear 
for a set of orthogonal set of projection operators and post-selected states.
 
Here we would like to re-emphasize the remark around Eq. (\ref{ABL_formula}) that the concept of weak value gives a physical basis of counter-factual arguments 
in Hardy's setting. The apparent paradox comes from the omission of the non-local possibility $I_pO_e+O_pI_e$ and the non-orthogonality of $O_p(I_e+O_e)$ 
and $(I_p+O_p)O_e$.
Hopefully the weak value is helpful to resolve quantum paradoxes which come from counter-factual arguments in more general settings.
\section{Spin Case} \label{Sec_spin}
So far we have mainly studied the projection operators as observables, which are relevant in  most of the optical experiments. In this section we study 
the non-commuting variables in the simple spin $1/2$ case. In contrast to the preceding case, we will encounter complex weak values and a new consistency 
condition. 

We choose the Pauli operators $\sigma_x, \sigma_y,$ and $\sigma_z$ as the three non-commuting observables. Let us start with an illustrative example: 
the pre-selected state is given by $\ket{\psi}=\ket{0_x}=\frac{1}{\sqrt{2}}(\ket{0}+\ket{1})$ and the post-selected 
states are $\bra{0_y}=\frac{1}{\sqrt{2}}(\bra{0}+i\bra{1}),\bra{1_y}=\frac{1}{\sqrt{2}}(\bra{0}-i\bra{1})$. The table for the weak values is
\begin{equation}
\begin{array}{c|ccc}
{\rm Weight} & 1/2 & 1/2 & \\
{\rm Post-selection} & \bra{0_y} & \bra{1_y} & {\rm Average} \\ \hline
\sigma_x & 1 & 1 & 1\\   
\sigma_y  & 1 & -1 & 0 \\  
\sigma_z & i & -i & 0\\ 
\end{array}.
\label{table4}
\end{equation}

Table (\ref{table4}) exhibits the fact that the operators $\sigma_x$ and $\sigma_y$ are orthogonal with respect to the pre-selected state $\ket{\psi}=\ket{0_x}$, since
$\bra{\psi}\sigma_x\sigma_y\ket{\psi}=i\bra{0_x}\sigma_z\ket{0_x}=0$. From the discussion of Sect.~\ref{Sec_formal}, we see that the weak vectors of $\sigma_x$ and 
$\sigma_y$ are real and orthogonal to each other, while that of $\sigma_z$ is complex.

We can also compose the non-orthogonal positive weak value vectors, e.g., as
\begin{equation}
\begin{array}{c|ccc}
{\rm Weight} & 1/2 & 1/2 & \\
{\rm Post-selection} & \bra{0_y} & \bra{1_y} & {\rm Average} \\ \hline
\sigma _x & 1 & 1 &  1 \\   
\sigma _x+\sigma _y & 2 & 0 & 1
\end{array}.
\label{table5}
\end{equation}
We can clearly see that if the non-orthogonal vectors of weak values are positive vectors, then the
orthogonal operators with respect to the preselected states contain at least one negative weak value. 

We would like to see how strange weak values, which
are outside the range of eigenvalues, show up in the spin $1/2$ case for the more general pre-selected state $\ket{\psi}=\alpha\ket{0}+\beta\ket{1}$
where $\alpha$ and $\beta$ are assumed to be real to simplify the exposition.
The post-selected states are $\ket{0}$ and $\ket{1}$. We compute the weak values,
\begin{equation}
\begin{array}{c|ccc}
{\rm Weight} & \alpha ^2 & \beta ^2 & {\rm (Bloch\; vectors)}\\
{\rm Post-selection} & \bra{0} & \bra{1} & {\rm Average} \\ \hline
\sigma_x & \beta /\alpha & \alpha /\beta & 2\alpha \beta \\   
\sigma_y & -i \beta /\alpha & i \alpha / \beta & 0 \\  
\sigma_z & 1  & -1 & \alpha^2 - \beta^2\\ \hline 
{\rm Sum \ of \ squared} & 1 & 1 & 1
\end{array},
\label{table6}
\end{equation}
where $``{\rm Sum \ of \ squared}" := ({\, _{f}\wv{\sigma_x}{\psi}^{w}})^2+({\, _{f}\wv{\sigma_y}{\psi}^{w}})^2 +({_{f}\wv{\sigma_z}{\psi}^{w}})^2 =1$.

Of course, they are not orthogonal to each other. The weighted average gives the Bloch vector of the pre-selected state $\ket{\psi}$ as shown 
in Table (\ref{table6}). The consistency with the multiplication rules, 
$ \sigma_x \sigma_y = i \sigma_z$ etc. can be checked, as well as the ``Sum of squared" results.

Let us see the weak value for a pair of orthogonal projection operators, $P_{\pm}:=(1 \pm \sigma_x)/2$:
\begin{equation}
\begin{array}{c|cc}
{\rm Weight} & \alpha ^2 & \beta ^2 \\
{\rm Post-selection} & \bra{0} & \bra{1} \\ \hline
P_{+} & \frac{1+ \beta /\alpha}{2} & \frac{1+\alpha /\beta}{2} \\   
P_{-} & \frac{1-\beta /\alpha}{2} & \frac{1-\alpha /\beta}{2} \\ \hline  
{\rm Sum} & 1 & 1
\end{array}.
\label{table7}
\end{equation}

It is evident that at least one of the weak values is negative. For $\beta / \alpha >1$ the weak value of $P_{-}$ with the post-selection $\bra{0}$ is negative.
This also means the weak value of $ \sigma _x $ for the post-selection $\bra{0}$
is strange, i.e., more than unity and thus outside the range of the eigenvalue spectrum discussed in Sect.~\ref{Sec_neg}.

The appearance of strange values can also be seen without
explicitly calculating the weak values.  Note that $({_{f}\wv{\sigma _x}{\psi}^{w}})^2 
+({_{f}\wv{\sigma _y}{\psi}^{w}})^2 +({_{f}\wv{\sigma _ z}{\psi}^{w})}^2=1$ in general for an arbitrary post-selected state $\bra{f}$. 
This implies that $_{0}\wv{\sigma _x}{\psi}^{w}
{_{1}\wv{\sigma _x}{\psi}^{w}}+_{0}\wv{\sigma _y}{\psi}^{w}{_{1}\wv{\sigma _y}{\psi}^{w}}+_{0}\wv{\sigma _z}{\psi}^{w}
{_{1}\wv{\sigma _z}{\psi}^{w}}=1$. By a suitable rotation, we will have
$_{0}\wv{\sigma _x}{\psi}^{w}{_{1}\wv{\sigma _x}{\psi}^{w}}=1$. We can see either $_{0}\wv{\sigma _x}{\psi}^{w}$ or $_{1}\wv{\sigma _x}{\psi}^{w}$ 
is strange, i.e., larger than $1$. The ABL formula cam be confirmed using 
$|\bkt{0}{\psi}|^2|_{f}\wv{\ket{0_x}\bra{0_x}}{\psi}^{w}|^2= (\alpha+\beta)^2 / 4 =|\bkt{0}{0_x}|^2|\bkt{0_x}{\psi}|^2$.
The intermediate states are orthogonal and therefore distinguishable so that they can be identified by a projective measurement, i.e., factual.
\section{Summary} \label{Sec_con}
In this paper, we have developed a framework of weak value vectors and derived the two consistency conditions for them which enables us to interpret 
them as (complex) conditional probabilities for counter-factual processes in quantum paradoxes. We have shown that at least one of the components of 
the weak value vector becomes strange for counter-factual process. Two examples Hardy's paradox and the spin $1/2$ system were demonstrated. In the former, the 
choice of the pre-selected state is unique from the setting of the paradox and the appearance of the ``negative probability" is due to the counter-factuality. 
In the latter we have discussed the case of non-commutative operators to weakly observe.

We would like to emphasize that the weak value should be studied on its own right and believe that this is just the right quantity to describe a sequence 
of counter-factual phenomena in quantum mechanics by virtue of the ``probabilistic" interpretation explained in this paper.
\section*{Acknowledgments}
The authors would like to acknowledge the useful comments of Professor Masanao Ozawa and 
the stimulating discussions with Professor Holger F. Hofmann. 
The authors are supported by Global Center of Excellence Program ``Nanoscience and Quantum Physics" at Tokyo Institute of Technology.
YS is also supported by JSPS Research Fellowships for Young Scientists (Grant No. 21008624). 

\end{document}